\documentclass[aps,prb,showpacs,amsmath,twocolumn,amssymb,superscriptaddress,letterpaper]{revtex4}
\usepackage{times}
\usepackage{amsfonts}
\usepackage{mathrsfs}
\usepackage{graphicx}
\usepackage{dcolumn}
\usepackage{bm}
\usepackage{color}

\usepackage[colorlinks,bookmarks=false,citecolor=blue,linkcolor=red,urlcolor=blue]{hyperref}
\def\be{\begin{equation}}       \def\ee{\end{equation}}
\def\bea{\begin{eqnarray}}      \def\eea{\end{eqnarray}}

\begin{document}
\title{Magnetisms in $p$-type monolayer gallium chalcogenides (GaSe, GaS)}
\author{Xianxin Wu}\email{xxwu@iphy.ac.cn}
\affiliation{ Institute of Physics, Chinese Academy of Sciences,
Beijing 100190, China} \affiliation{ Department of Physics and
Center of Theoretical and Computational
Physics,\\
The University of Hong Kong, Hong Kong, China}

\author{Xia Dai}
\affiliation{ Institute of Physics, Chinese Academy of Sciences,
Beijing 100190, China}

\author{Hongyi Yu}
\affiliation{ Department of Physics and Center of Theoretical and
Computational
Physics,\\
The University of Hong Kong, Hong Kong, China}

\author{Heng Fan }   \affiliation{ Institute of Physics, Chinese Academy of Sciences,
Beijing 100190, China}

\author{Jiangping Hu  }\affiliation{
Institute of Physics, Chinese Academy of Sciences, Beijing 100190,
China}\affiliation{Department of Physics, Purdue
University, West Lafayette, Indiana 47907, USA}

\author{Wang Yao}  \affiliation{ Department of Physics and Center of Theoretical and Computational
Physics,\\
The University of Hong Kong, Hong Kong, China}

\date{\today}

\begin{abstract}
Magnetisms in $p$-type monolayer GaX (X=S,Se) is investigated by
performing density-functional calculations. Due to the large density
of states near the valence band edge, these monolayer semiconductors are
ferromagnetic within a small range of hole doping. The intrinsic Ga
vacancies can promote local magnetic moment while Se vacancies
cannot. Magnetic coupling between vacancy-induced local moments is
ferromagnetic and surprisingly long-range. The results indicate that
magnetization can be induced by hole doping and can be
tuned by controlled defect generation.
\end{abstract}


\maketitle

\section{Introduction}
The researches of two-dimensional (2D) materials have attracted extensive interest recently. A varieties of 2D crystals have been discovered which exhibit remarkable properties. These include graphene\cite{Novoselov2004,Novoselov2005,Zhang2005,Geim2007}, a gapless semiconductor, single-layer boron nitride\cite{Kim2012,Tusche2007}, a wide-gap insulator, and
monolayer transition metal dichalcogenides, a class of direct band-gap semiconductor with excotic spin and pseudospin physics\cite{Xiao2012,Wang2012,Splendiani2010,Mak2010}.
Semiconducting 2D crystals are of interest as the next generation host materials for future electronic applications exploiting internal degrees of freedom of carriers such as spin in addition to the charge\cite{Xu2014}.
An indispensible element of spin based electronics is magnetism which can play the role of non-volatile memory. Realizing robust magnetic semiconductor has been a long term effort in the exploration of semicondutor-based spintronics\cite{Wolf2001}. The conventional approach to introduce magnetism in semiconductors is by doping transition metal elements\cite{Ohon1998,Dietl2000,Jungwirth2006}. Recently,
unexpected high-temperature ferromagnetism have been
reported in $d^0$ systems\cite{Monnier2001,Elfimov2002,Venkatesan2004,Pan2007,Dev2008,Peng2009}, which do not involve the partially
occupied $d$ or $f$ orbitals. In
carbon structures, first-principles calculations have also shown that magnetism can be induced by nonmagnetic defects\cite{Kusakabe2003,Kim2003,Chan2004,Andriotis2006}. While the origin of magnetism in these $d^0$ systems is not well understood, the large density of states at the band edge is considered to be the key for several materials\cite{Peng2009}, which implies a new possibility for searching spintronic materials. In 2D crystals like BN, graphene and MoS$_2$, first-principle calculations have also predicted that local magnetic moment can be induced by nonmagnetic impurities or vacancies\cite{Liu2007,Yazyev2007,Zhang2007,Zhou2013}.

\begin{figure}[t]
\centerline{\includegraphics[height=8.5 cm]{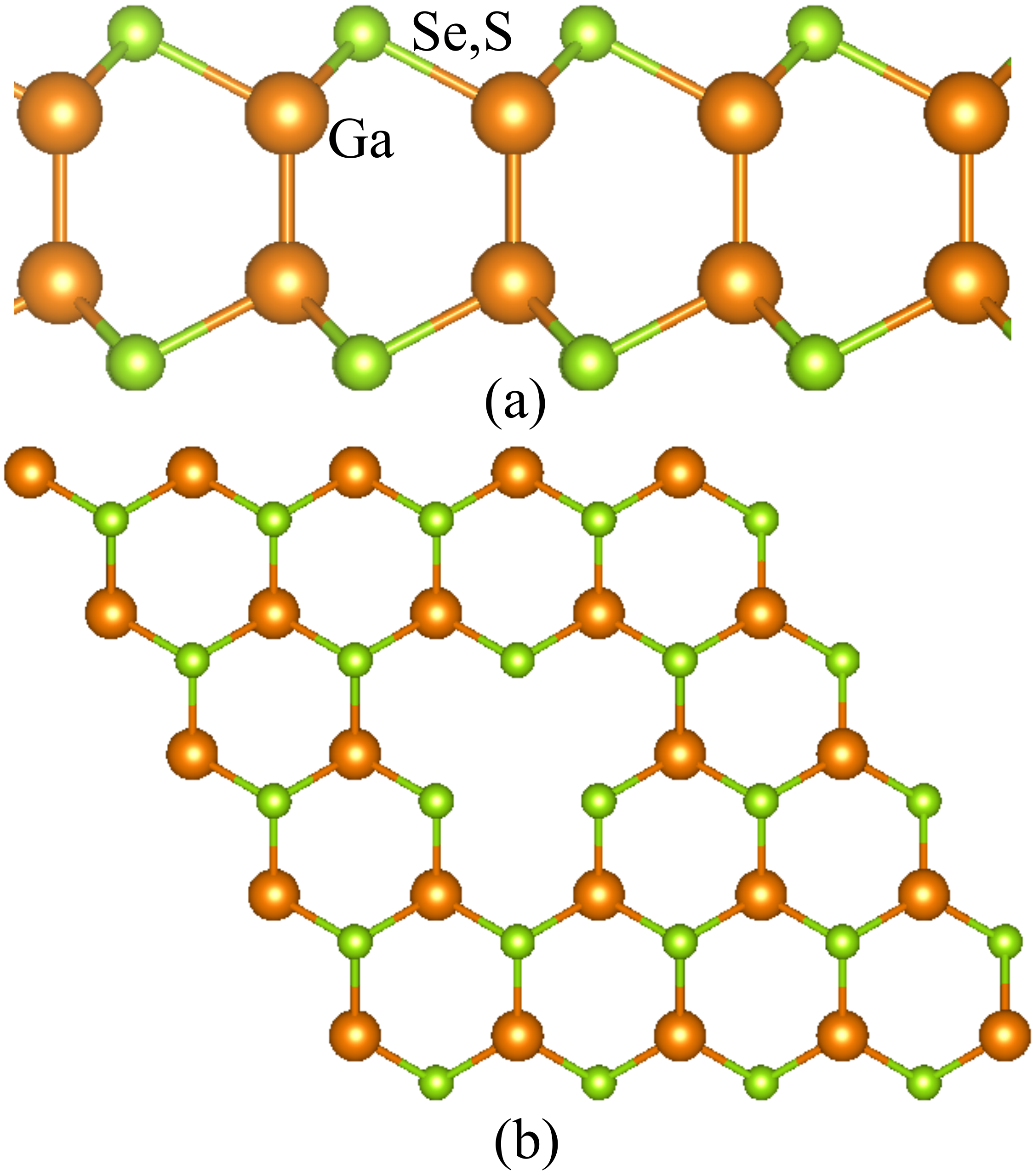}}
 \caption{ (Color online) Monolayer GaSe: (a) Side view, (b) Top view with two Ga vacancies.
 \label{GaSemono} }
\end{figure}

GaSe and GaS are stable layered semiconductors, formed by vertically stacked
X-Ga-Ga-X sheets held together by van der Waals interactions, as
shown in Fig.~\ref{GaSemono}(a). They have attracted considerable
interest because of their remarkable nonlinear optical
properties\cite{Segura1997,Nusse1997,Kato2011}. Very recently,
ultrathin GaS and GaSe nanosheets have been synthesized and
experimentally studied\cite{Hu2012,Late2012}. Zolyomi \emph{et al.} have
performed first-principles calculations of GaX monolayers and found
a sombrero dispersion near the top of the valence
band\cite{Zolyomi2013}, which gives rise to a large density
of states at the valence band edge (VBE). Therefore, monolayer GaX
can be a promising system to explore $d^0$ ferromagnetism in the 2D limit.

In this manuscript, we investigate the magnetism in hole-doped GaX monolayers using density functional theory calculations. We find that GaX monolayer becomes ferromagnetic with a small range of hole doping. In this range, the polarization energies first increases then decreases to zero. Under Se-rich condition, a Ga vacancy is more favorable, rendering the
GaSe intrinsic $p$ type. A S vacancy can be spontaneously introduced
and makes GaS intrinsic $n$ type. These are in agreement with the
experiment\cite{Late2012}.
The intrinsic Ga vacancies can promote the formation of
local magnetic moments. We also find that the magnetic coupling between these defects is
ferromagnetic and surprisingly long-ranged.

\section{Electronic structures and hole-induced magnetism of monolayer GaSe and GaS }\label{S1}

\subsection{Computation method}
 Our density functional theory (DFT) calculations employ the
projector augmented wave (PAW) method encoded in Vienna \emph{ab
initio} simulation package(VASP)
\cite{Kresse1993,Kresse1996,Kresse1996B}, and the
generalized-gradient approximation (GGA) for the exchange
correlation functional \cite{Perdew1996} is used. Throughout this
work, the cutoff energy of 400 eV is taken for expanding the wave
functions into plane-wave basis. The calculated lattice constants for
monolayer GaSe and GaS are 3.817 and 3.627 \AA, respectively. A
$12\times12\times1$ $\Gamma$ centered k-point grids are used to
sample the Brillouin zone for the primitive cell. In the hole doping calculation, $27\times27\times1$ k-point grids are adopted.  For the vacancy
calculation, we adopt $4\times4$ supercell to simulate the system
with defects, which is large enough to avoid the interaction between
the defects. A set of $3\times3\times1$ $\Gamma$ centered k-points
is used for the defect calculation. A 15 \AA vacuum layer is used in
our calculation to avoid interaction between slab. The convergence
for energy is chosen as $10^{-5}$ eV between two steps and the
maximum Hellmann-Feynman force acting on each atom is less than 0.02
eV/\AA upon ionic relaxation for decfect calculation( 0.01 eV/\AA
for primitive cell GaX).

The formation energy of neutral vacancies in monolayer GaSe or GaS,
$\Delta E_f$, is defined as,
\begin{equation}
\Delta E_f=E_t(nV_Y)-E_t+n\mu_Y
\end{equation}
where $E_t(nV_Y)$ is the total energy of monolayer GaSe or GaS with
$n$ Y (Y=Ga, Se, S) vacancies. $E_t$ is the energy of pristine
monolayer GaSe or GaS and $\mu_X$ refers to the chemical potential
of detached X atom. $\mu_Y$ is obviously environment dependent. We
consider two cases: Ga-rich and Se-rich or S-rich. In Ga-rich
environment, $\mu_{Ga}$ is calculated from orthorhombic Ga crystal
and $\mu_{Se/S}=\mu_{GaSe/GaS}-\mu_{Ga}$, where $\mu_{GaSe}$
($\mu_{GaS}$) is the chemical potential of a GaSe (GaS) unit in
2H-GaSe (2H-GaS) crystal. In Se-rich or S-rich environment,
$\mu_{Se}$ ($\mu_S$) is calculated from hexagonal Se (orthorhombic
S) crystal and $\mu_{Ga}=\mu_{GaSe/GaS}-\mu_{Se/S}$.

\subsection{Electronic structures and hole-induced magnetism for Monolayer GaSe and GaS}

\begin{figure*}[t]
\centerline{\includegraphics[height=7
cm]{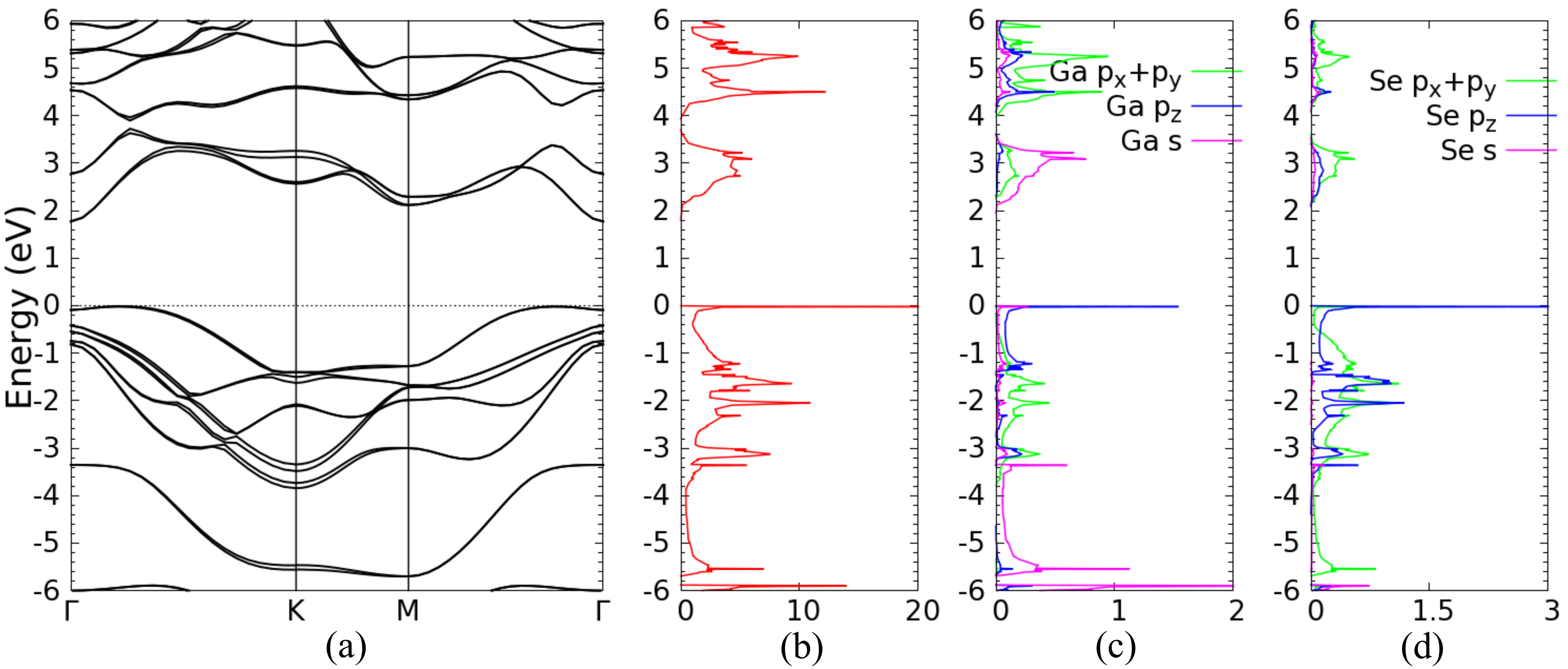}}
 \caption{ (Color online) The electronic structure of monolayer GaSe with spin orbital coupling. (a) bandstructure of monolayer GaSe. (b) The total DOS; (c) and (d) are the orbital projected density of states for Ga and Se.
 \label{GaSebanddos} }
\end{figure*}

\begin{figure*}[t]
\centerline{\includegraphics[height=7
cm]{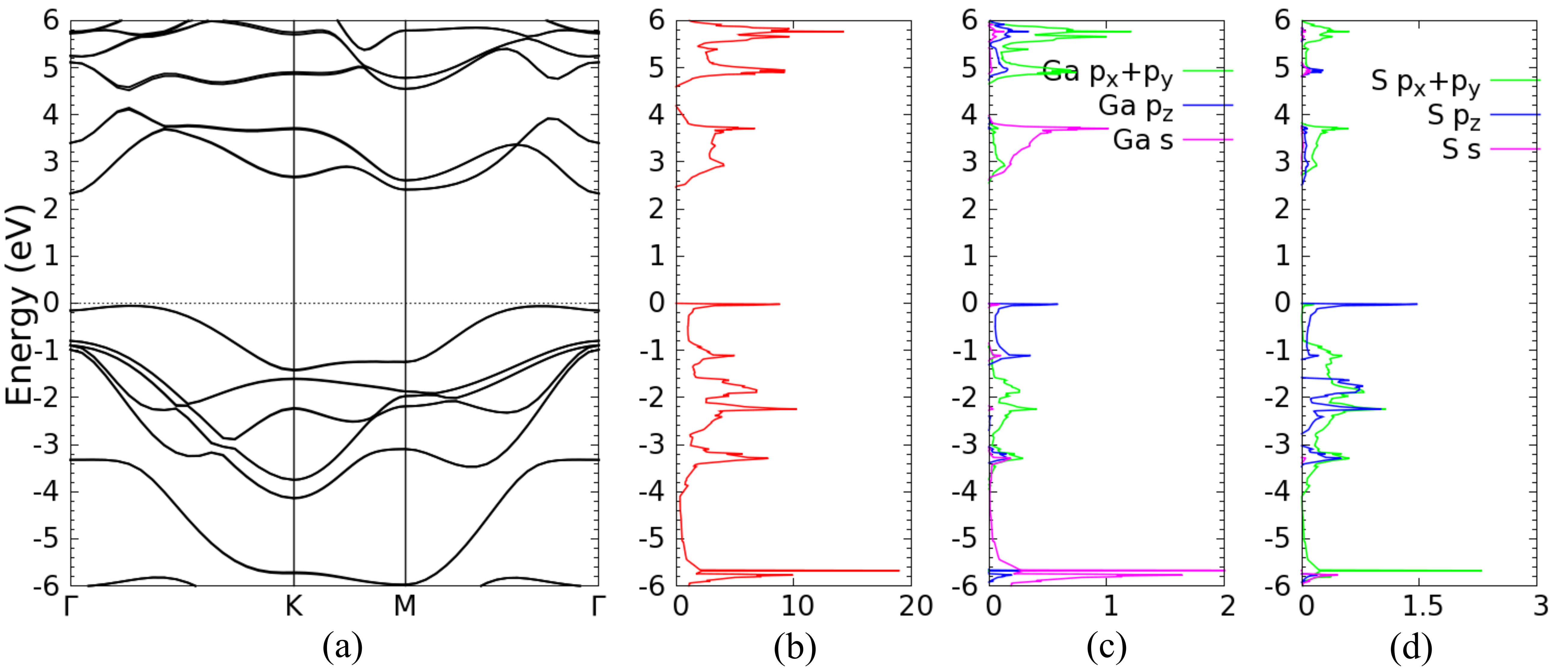}}
 \caption{ (Color online) The electronic structure of monolayer GaS with spin orbital coupling. (a) bandstructure of monolayer GaS. (b) The total DOS; (c) and (d) are the orbital projected density of states for Ga and S.
 \label{GaSbanddos} }
\end{figure*}

The band structures of monolayer GaSe and GaS with spin orbit
coupling are presented in Fig.~\ref{GaSebanddos} and
Fig~\ref{GaSbanddos}, respectively. Both monolayer GaSe and GaS are
semiconductors with indirect bandgaps, which are just slightly lower
than the direct bandgaps. The bandgaps of monolayer GaSe and GaS are
2.1 eV and 2.5 eV, repsectively. The bottom of the conduction band
is mainly attributed to Ga $s$ orbital, hybridized with $p_x$ and
$p_y$ orbitals of X (X=Se, S).  The top of the valence band is
primarily attributed to X $p_z$ orbitals, hybridized strongly with
Ga $p_z$. Due this orbital character, the band near $\Gamma$ point
shows rather flat dispersion, which induces a Van Hove singularity
near the valence band edge. The density of states (DOS) of GaSe near
VBE is twice of that of GaS, indicating that the band of GaSe is
flatter. The spin orbital coupling yields the valence band splitting
by 10 meV at the maximum point in $\Gamma$-M direction for GaSe
while this splitting for GaS is 4 meV.

\begin{figure}[t]
\centerline{\includegraphics[height=10 cm]{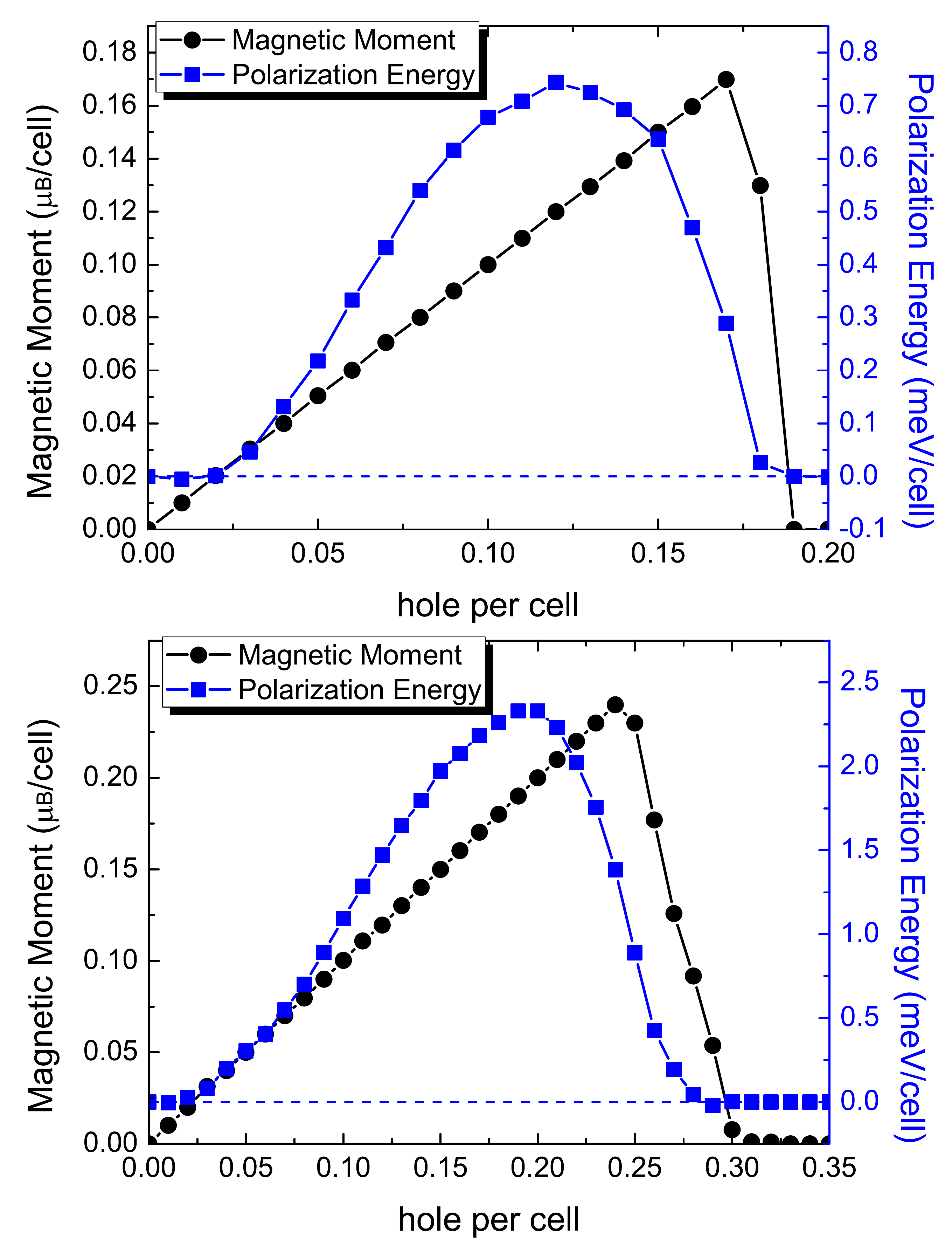}}
 \caption{ (Color online) The top panel and bottom panel show Magnetic moments and polarization energy of hole-doped monolayer GaSe and GaS, repectively.
 \label{GaSehole} }
\end{figure}

In the band-picture model, spontaneous ferromagnetism appears when
the relative exchange interaction is larger than the loss in kinetic
energy, that is, when it is satisfies the "Stoner Criterion":
$D(E_f)J>1$, where $D(E_f)$ is the DOS at the Fermi energy $E_f$ and
$J$ is the strength of the exchange
interaction\cite{Stoner1938,Peng2009}. As the DOS near VBE is rather
large, we can dope hole into the system to increase $D(E_f)$. We
check the stability of magnetization by calculating the polarization
energy $E_p$, which is defined as the difference between the
nonspin-polarized and spin-polarized states. From
Fig.\ref{GaSehole}, we find that the ground state of the monolayer
GaSe is ferromagnetic when the hole concentration is in the range
from 0.02 to 0.18 per unit cell. While, the monolayer GaS becomes
ferromagnetic in a wider range of  hole concertration and the maximum hole
concentration is 0.28 per unit cell from Fig~\ref{GaSehole}.
When the hole concentration is not in this range, $D(E_f)$ is not
large enough and the "Stoner Criterion" is not satisfied, thus the
system is nonmagnetic. The critical hole concentration for GaSe
(0.18 hole per cell) corresponds to 1.24$\times 10^{14}$ cm$^{-2}$.
The low hole concentration in the system can be easily achieved by
electric carrier doping, which has been used to induce
superconductivity in MoS$_2$\cite{Taniguchi2012,Ye2012}. The
polarization energies in both cases first increase then decrease to
zero with the increasing hole doping. The magnetic moment shows a
linear relationship with hole concentration when the system is
ferromagnetic. At optimal doping, polarization energies are 0.74 meV
and 2.33 meV per cell for Ga$_2$Se$_2$ and Ga$_2$S$_2$,
respectively. The corresponding magnetic moments are 0.12 and 0.20
$\mu_B$ per cell.

\section{The vacancy in Monolayer GaSe and GaS}
According to Ref.~\onlinecite{Late2012}, the single-sheet GaS and
GaSe show typitcal n-type and p-type conductance, respectively.
These may be the consequence of intrinsic vacancies. The calculated formation
energies of vacancies in both cases are shown in
Table \ref{formationE}. In Se rich condition, the formation energy
of a Ga vacancy is slightly less than that of a Se vacancy. Thus, a
Ga vacancy is more favorable. This may explain the observed p-type
conductance in GaSe. However, a Se vacancy is more favorable under
Ga-rich condition. In GaS, the formation energies of a S vacancy are
always negative under both conditions, indicating that S vacancies
can be spontaneously introduced. Therefore, GaS is expected to be an
intrinsic $n$ type, which is in agreement with experiment. The
formation energy of a Ga vacancy in GaSe is less than that in GaS.
In order to get magnetization, we need to dope hole into above
systems, thus only Ga vacancies are discussed in the following.

\begin{table}[bt]
\caption{\label{formationE}%
The formation energies of vacancies in monolayer GaSe and GaS under Ga-rich and Se-rich or S-rich conditions. V$_{2Ga}$} denotes the two bonded Ga vacancy.
\begin{ruledtabular}
\begin{tabular}{ccccccc}
&\multicolumn{3}{c}{GaSe} & \multicolumn{3}{c}{GaS}\\
Environment & $V_{Ga}$ & $V_{2Ga}$  & $V_{Se}$  & $V_{Ga}$  & $V_{2Ga}$  & $V_{S}$ \\
 \colrule
Ga rich & 2.97 & 5.72 & 1.53 & 4.29 & 7.35 & -3.03 \\
Se or S rich  & 1.81 & 3.40 & 2.69 & 2.93 & 4.64 & -0.32\\
\end{tabular}
\end{ruledtabular}
\end{table}

\subsection{one Ga vacancy}
\begin{figure}[t]
\centerline{\includegraphics[height=8.5 cm]{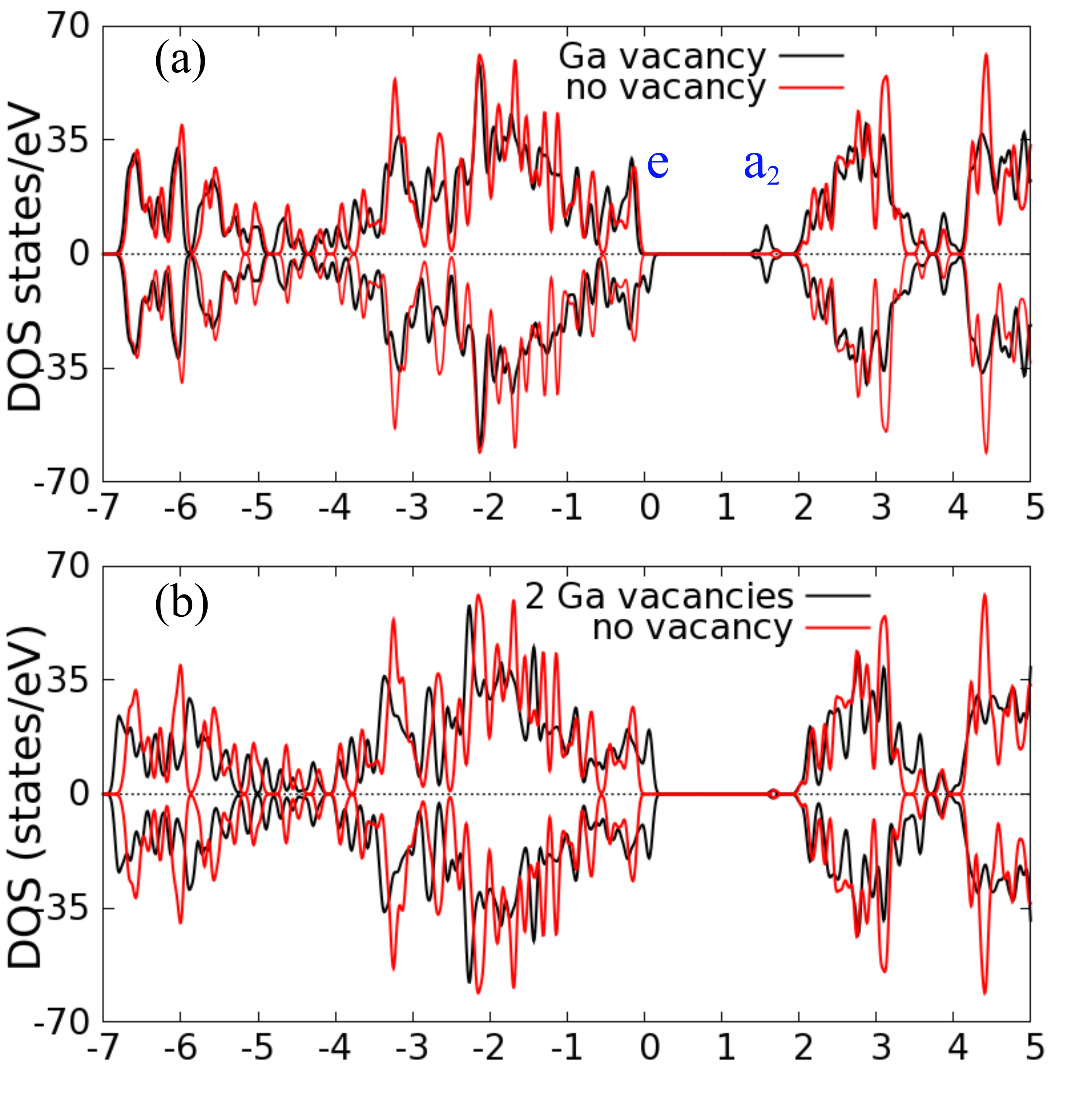}}
 \caption{ (Color online) Spin-resolved density of states of one and two Ga vacancies in 4$\times$4 supercell containing 64 atoms. The top panel shows DOS for the case of one vacancy while the the bottom panel shows DOS for the case of two vacancies after relaxation. The positive and negative value of DOS denote the majority spin states and the minority spin states, respectively.
 \label{Ga1dos} }
\end{figure}

\begin{figure}[t]
\centerline{\includegraphics[height=4.5 cm]{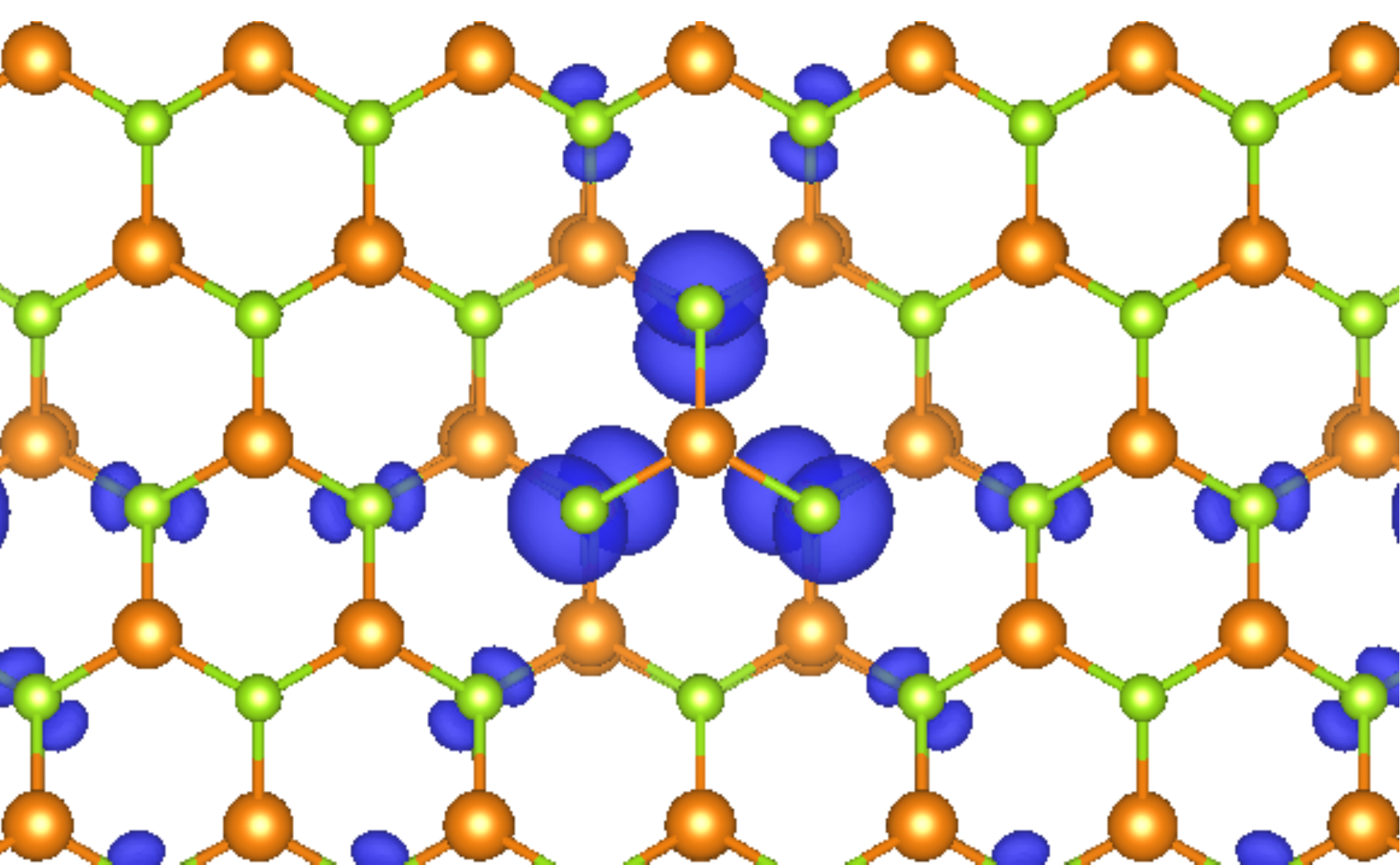}}
 \caption{ (Color online) Isosurface spin-density plot ($\rho=\rho_{\uparrow}-\rho_{\downarrow}$) for one Ga vacancy in GaSe/GaS monolayer system.
 \label{Ga1mag} }
\end{figure}

After relaxation, Se and S atoms surrounding the neutral Ga vacancy
move outward while the Ga under the vacancy moves downward.  Thus,
the Se-Ga bond lengths in the top and the bottom layer shrink
from 2.50 \AA to 2.41 \AA and 2.37 \AA in monolayer GaSe,
respectively. In monolayer GaS, the bond lengths of Ga-S in top and
the bottom layer shrinks from 2.35 \AA to 2.25 \AA and 2.22 \AA,
respectively. Fig.\ref{Ga1dos}(a) shows the spin-resolved density of
states (DOS) of the 4$\times$4 supercell containing one Ga vacancy. A
local moment of 1.0 $\mu_B$ is formed due to the spin polarization
in both cases. The magnetic moments mainly locate at the three Se or
S atoms surrounding the Ga vacancy, shown in Fig.\ref{Ga1mag}.

The formation of the local moment by Ga vacancy can be
understood: The point group of monolayer GaX (X=Se, S) with a Ga
vacancy is $C_{3v}$. A Ga vacancy leaves five unpaired electrons on
the X and Ga atoms surrounding the vacancy. Under the influence of
the crystal field, defect states associated with Ga vacancy are
split into two singlet $a_1$, $a_2$ and a doublet $e$. The fully
occupied $a_1$ state lies below the fermi level while $a_2$ is
empty. Due to spin polarization, the $e$ state splits into spin-up
$e_{\uparrow\uparrow}$ and spin-down $e_{\downarrow}$. Therefore, a
Ga vacancy should result in a net local moment of 1 $\mu_B$. The
polarization energies $E_p$ are about 0.03 eV per defect site in
GaSe and GaS.

\subsection{Two Ga vacancies}
With the vacancy of two bonded Ga , the mirror symmetry is restored
in the system.  The total magnetic moment, mainly localized
at the Se atoms surrounding the vacancy, is 3.95 $\mu_B$ without
relaxation. However, the magnetic moment vanishes after relaxation
due to the movement of Se atoms surrounding the Ga vacancy. The
length of Ga-Se bonds in both layers shrinks to 2.38 \AA. The spin-resolved density of
DOS of the 4$\times$4 supercell containing two Ga vacancies shown in Fig.\ref{Ga1dos}(b).
 A local moment of 1.0 $\mu_B$ is formed due to the spin polarization
in both cases.

Two Ga vacancies leave eight unpaired electrons on the surrounding
Se atoms. Before relaxation, under the crystal field defect states
are split into four singlets $a_1$, $a'_1$, $a_2$, $a'_2$ and two
doublets $e_1$, $e_2$. $a_1$ and $a'_1$ are fully occupied while
$a_2$ and $a'_2$ are empty. The remaining four electrons occupied
$e_1^{\uparrow\uparrow}$ and $e_2^{\uparrow\uparrow}$ due to large
spin splitting. Thus, a net moment of 4 $\mu_B$ is formed. However,
the energy of the $e_2$ states increases after relaxation. The
energy of $e_2^{\uparrow}$ is higher than that $e_1^{\downarrow}$.
Therefore, the $e_1$ states are fully occupied, which results in a
nonmagnetic state after relaxation.

\section{The exchange coupling between vacancies }

To investigate the magnetic coupling between these hole-induced
magnetic moments, we employ the $8\times4$ supercell with two
vacancies in each subsupercell of $4\times4$. Two stable magnetic
configurations (ferromagnetic and antiferromagnetic) can be obtained
depending on the initial moments in the calculation. The Heisenberg
type of spin coupling is: $H=-\sum_{<ij>} J_{ij}\textbf{S}_i\cdot
\textbf{S}_j$. Considering nearest-neighbor interactions, the
magnetic energy of the FM state is: $E_{FM}=-6JS^2$. For the case of
AFM states, four nearest-neighbors have spins parallel and two have
spins antiparallel. Thus, the magnetic energy is : $E_{AFM}=-2JS^2$.
The energy difference of AFM and FM states is
$E_M=E_{AFM}-E_{FM}=4JS^2$.  Table \ref{magneticE} shows the energy
difference $E_M$ and exchange coupling parameter $J$ for monolayer
GaSe and GaS. In FM states, the total magnetic moment is always 2
$\mu_B$ for both cases. From Table \ref{magneticE}, we find that FM
states has a lower energy than the AFM states.

The observed FM state can be explained by kinetic exchange
mechanism\cite{Dev2008}. For the isolated Ga vacancy, the majority
spin ($e^{\uparrow\uparrow}$) is fully occupied, where the minority
spin ($e^{\downarrow}$) is partially occupied. Therefore, the
parallel spin alignment allows for the virtual hopping between the
two defects states, which can lower the kinetic energy of the
system. This hoping, however, is not allowed if the spin alignment
is antiparallel.

We estimate the Curie temperature ($T_c$) based on the mean-field theory and Heisenberg model using the equation,
\begin{equation}
 k_BT_C=\frac{2}{3}J_0,
 \end{equation}
where $J_0$ is the onsite exchange parameter reflecting the exchange
field created by all the neighboring magnetic moments. The estimated
$T_C$ for monolayer GaSe and GaS are 50 and 26 K, respectively. The
exchange coupling between local moment in GaSe is almost twice of
that in GaS. It indicates the impurity state in GaS is more
localized than that in GaSe.

\begin{table}[bt]
\caption{\label{magneticE}%
The energy difference between the FM and AFM state $\Delta E$=$E_{AFM}$-$E_{FM}$ for four vacancy configurations.
$d$ is the distance between the two vacancies. }

%
%

\begin{ruledtabular}
\begin{tabular}{ccccc}

System & $E_M$ (meV) & J (meV) & $T_C$ (K)\\
 \colrule

$GaSe $ & 26.1 & 6.5 & 50 \\
$GaS $ & 13.4 & 3.4 & 26
\end{tabular}
\end{ruledtabular}
\end{table}

\section{Conclusion}
In summary, we perform DFT calculations to investigate the
electronic structure of monolayer GaX. We find that monolayer GaSe(GaS)
is a semiconductor with an indirect bandgap of 2.1(2.5) eV and there is a
Van Hove singularity near the valence band edge. The monolayer GaX
becomes ferromagnetic with small hole doping, which may be achieved
by electric carrier doping in experiment. Under Se-rich condition,
GaSe is intrinsic $p$ type induced by Ga vacancies. For GaS, a S
vacancy can be spontaneously introduced, rendering GaS $n$ type. Ga
vacancies can induce local moment around this defect. The coupling
between the states of two Ga vacancies is ferromagnetic and
extremely long-range.

\section{Acknowledgments}

We thank H. M. Weng for extremely useful discussion.  The work is supported by "973" program (Grant No.  2010CB922904 and No. 2012CV821400), the National Science Foundation of China (Grant No. NSFC-1190024, 11175248 and 11104339), and the Research Grant Council of Hong Kong SAR (Grant No. HKU706412P).


{\it{Note added}}--After this work is completed, we became aware of the paper\cite{Cao2012}, which addresses the magnetism and half-metallicity in hole-doped monolayer GaSe.


\begin{thebibliography}{multiferroics}
%
\bibitem{Novoselov2004} K. S. Novoselov et al., Science {\bf 306}, 666 (2004).
\bibitem{Novoselov2005} K. S. Novoselov et al., Nature (London) {\bf 438}, 197 (2005).
\bibitem{Zhang2005} Y. Zhang et al., Nature (London) {\bf 438}, 201 (2005).
\bibitem{Geim2007} A. K. Geim and K. S. Novoselov, Nature Mater. {\bf 6} , 183 (2007).

\bibitem{Kim2012} K. K. Kim, Y. Shi, M. Hofmann, D. Nezich, J.?F. Rodriguez-Nieva, M.
Dresselhaus, T. Palacios, and J. Kong, Nano Lett. {\bf 12}, 161
(2012).
\bibitem{Tusche2007} C. Tusche, H. L. Meyerheim, and J. Kirschner, Phys. Rev.
Lett. {\bf 99}, 026102 (2007).


\bibitem{Xiao2012} D. Xiao, G.-B. Liu, W. Feng, X. Xu,
and W. Yao, Phys. Rev. Lett. {\bf 108}, 196802 (2012).
\bibitem{Wang2012}  Q. H. Wang, K. Kalantar-Zadeh, A. Kis, J. N. Coleman and M. S. Strano, Nature Nanotech., {\bf 7}, 699 (2012).
\bibitem{Splendiani2010}  A. Splendiani, L. Sun, Y. Zhang, T. Li, J. Kim, C.-Y. Chim, G. Galli and F. Wang, Nano Lett., {\bf 10}, 1271(2010).
 \bibitem{Mak2010}  K. F. Mak, C. Lee, J. Hone, J. Shan and T. F. Heinz, Phys. Rev. Lett., {\bf 105}, 136805(2010).



\bibitem{Xu2014} X. Xu, W. Yao, D. Xiao and T. F. Heinz, Nature Phys., {\bf 10}, 343(2014).


\bibitem{Coleman2011} J. N. Coleman, M. Lotya, A. O¡¯Neill, S. D. Bergin, P. J. King, U.
Khan, K. Young, A. Gaucher, S. De, R. J. Smith, I. V. Shvets, S. K.
Arora, G. Stanton, H.-Y. Kim, K. Lee, G. T. Kim, G. S. Duesberg, T.
Hallam, J. J. Boland, J. J. Wang, J. F. Donegan, J. C. Grunlan, G.
Moriarty, A. Shmeliov, R. J. Nicholls, J. M. Perkins, E. M.
Grieveson, K. Theuwissen, D. W. McComb, P. D. Nellist, and V.
Nicolosi, Science {\bf 331}, 568 (2011).

\bibitem{Wolf2001} S. A. Wolf, D. D. Awschalom, R. A. Buhrman, J. M. Daughton, S. von Moln¨¢r, M. L. Roukes, A. Y. Chtchelkanova, and D. M. Treger, Science {\bf 294}, 1448 (2001).
\bibitem{Ohon1998} H. Ohno, Science {\bf 281}, 951 (1998).
\bibitem{Dietl2000} T. Dietl, H. Ohno, F. Matsukura, J. Cibert, and D. Ferrand,
Science {\bf 287} , 1019 (2000).
\bibitem{Jungwirth2006} T. Jungwirth, J. Sinova, J. Masek , J. Kucera, and A. H.
MacDonald, Rev. Mod. Phys. {\bf 78}, 809 (2006).

\bibitem{Monnier2001} R. Monnier and B. Delley, Phys. Rev. Lett. {\bf 87}, 157204 (2001)
\bibitem{Elfimov2002} I. S. Elfimov, S. Yunoki, and G. A. Sawatzky, Phys. Rev. Lett. {\bf 89}, 216403 (2002).
\bibitem{Venkatesan2004} M. Venkatesan, C. B. Fitzgerald, and J. M. D. Coey, Nature (London) {\bf 430}, 630 (2004).
\bibitem{Pan2007} H. Pan, J. B. Yi, L. Shen, R. Q. Wu, J. H. Yang, J. Y. Lin, Y. P. Feng, J. Ding L. H. Van, and J. H. Yin, Phys. Rev. Lett. {\bf 99}, 127201 (2007).

\bibitem{Dev2008} P. Dev, Y. Xue and P. H. Zhang, Phys. Rev. Lett. {\bf 100},
117204 (2008).
\bibitem{Peng2009} H. W. Peng, H. J. Xiang, S. H. Wei, S. S. Li J. B. Xia and J. B. Li, Phys. Rev. Lett. {\bf 102},
017201 (2009).



\bibitem{Kusakabe2003} K. Kusakabe and M. Maruyama, Phys. Rev. B {\bf 67}, 092406 (2003).
\bibitem{Kim2003} Y.-H. Kim, J. Choi, K. J. Chang, and D. Tomanek, Phys. Rev. B { \bf 68}, 125420 (2003).
\bibitem{Chan2004} J. A. Chan et al., Phys. Rev. B {\bf 70}, 041403 (2004).
\bibitem{Andriotis2006} A. N. Andriotis, R. M. Sheetz, E. Richter, and M. Menon
Europhys. Lett.72, 658 (2005).

\bibitem{Liu2007} R. F. Liu and C. Cheng, Phys. Rev. B {\bf 76}, 014405 (2007).
\bibitem{Yazyev2007} O. V. Yazyev and L. Helm, Phys. Rev. B {\bf 75}, 125408 (2007).
\bibitem{Zhang2007} Y. Zhang, S. Talapatra, S. Kar, R. Vajtai, S. K. Nayak, and P. M. Ajayan, Phys. Rev. Lett. {\bf 99}, 107201 (2007).

\bibitem{Zhou2013} Y. G. Zhou, P. Yang, H. Y. Zu, F. Gao and X. T. Zu, Phys. Chem. Chem. Phys., {\bf 15}, 10385 (2013).





\bibitem{Segura1997} A. Segura, J. Bouvier, M.V. Andres, F. J. Manjon, and V. Munoz. Phys. Rev. B, {\bf 56}, 4075 (1997).
\bibitem{Nusse1997} S. Nusse, P.H. Bolivar, H. Kurz, V. Klimov and F. Levy, Phys. Rev. B, {\bf 56}, 4578 (1997).
\bibitem{Kato2011} K. Kato, N. Umemura, Optics Letters, {\bf 36}, 746 (2011).

\bibitem{Hu2012} P.A. Hu, Z.Z. Wen, L.F. Wang, P.H. Tan, K. Xiao, ACS Nano, {\bf 6}, 5988 (2012).
\bibitem{Late2012} D. J. Late, B. Liu, J.J. Luo, A.M. Yan, H.S.S.R. Matte, M. Grayson, C.N.R. Rao, V.P. Dravid, Advanced Materials, {\bf 24} 3549 (2012).

\bibitem{Zolyomi2013} V. Zolyomi, N. D. Drummond and V. I. Falko, Phys. Rev. B {\bf 87}, 195403.

\bibitem{Kresse1993} G. Kresse and J. Hafner, Phys. Rev. B {\bf 47}, 558 (1993).
\bibitem{Kresse1996} G. Kresse and J. Furthmuller, Comput. Mater. Sci. {\bf 6}, 15 (1996).
\bibitem{Kresse1996B} G. Kresse and J. Furthmuller, Phys. Rev. B {\bf 54}, 11169 (1996).


\bibitem{Perdew1996} J. P. Perdew, K. Burke, and M. Ernzerhof, Phys. Rev. Lett. {\bf 77},
3865 (1996).


\bibitem{Stoner1938} E. C. Stoner, Proc. R. Soc. A {\bf 165}, 372 (1938); {\bf 169}, 339 (1939).


\bibitem{Taniguchi2012} K. Taniguchi, A. Matsumoto, H. Shimotani and H. Takagi,
Appl. Phys. Lett. {\bf 101}, 042603 (2012).
\bibitem{Ye2012} J. T. Ye, Y. J. Zhang, R. Akashi, M. S. Bahramy, R. Arita and
Y. Iwasa, Science {\bf 338}, 1193(2012).


\bibitem{Cao2012} T. Cao, Z. L. Li, S. G. Louie,arxiv:1409.4112.

\end{thebibliography}
\end{document}